%% file: main.tex
\documentclass[runningheads]{llncs}

\usepackage{verbatim}
\usepackage{multirow}
\usepackage{wrapfig}
\usepackage[breaklinks,hidelinks]{hyperref}

\usepackage[normalem]{ulem}

\usepackage{tikz}
\usepackage{listings,color}
\usetikzlibrary{shadows,shapes,shapes.geometric,arrows}
\usepackage{pgfplots}
\pgfplotsset{compat=1.10}

\definecolor{dkgreen}{rgb}{0,0.6,0}
\definecolor{forestgreen}{RGB}{0,100,50}
\definecolor{redd}{RGB}{76,0,153}
\definecolor{english}{rgb}{0.0, 0.26, 0.15}
\definecolor{coolblack}{rgb}{0.0, 0.18, 0.39}
\definecolor{blue-violet}{rgb}{0.54, 0.17, 0.89}

\lstdefinestyle{customjava}
{language=java,
keywordstyle=\color{blue-violet},
basicstyle=\ttfamily\scriptsize,
morekeywords={String, var, require, msg, keccak256, now, function, continue, else},
mathescape=true,
escapeinside={/*@}{@*/},
commentstyle=\color{english},   
keywordstyle=[2]{\color{red}},
}

\lstdefinestyle{customjavaexample}
{language=java,
keywordstyle=\color{blue-violet},
basicstyle=\ttfamily\scriptsize,
morekeywords={String, var, require, msg, keccak256, now, function, continue, else},
mathescape=true,
escapeinside={/*@}{@*/},
commentstyle=\color{english},   
keywordstyle=[2]{\color{red}},
frame=single
}

\usepackage[textsize=tiny,textwidth=4cm]{todonotes}
\usepackage{amsmath}
\usepackage{amsfonts}
\usepackage{paralist}

\newcommand{\code}[1]{\textcolor{blue}{\texttt{#1}}}
\newcommand{\lbl}[1]{\textit{#1}}
\newcommand{\ind}{~~~}

\begin{document}
\setlength{\marginparwidth}{4cm}

\title{Tool Demonstration:\\
FSolidM for Designing\\Secure Ethereum Smart Contracts}
\titlerunning{FSolidM for Designing Secure Ethereum Smart Contracts}
\author{Anastasia Mavridou\inst{1} \and Aron Laszka\inst{2}}
\authorrunning{Anastasia Mavridou and Aron Laszka}
\institute{}
\institute{Vanderbilt University \and University of Houston}
\maketitle

 \begin{center}
 Accepted for publication in the proceedings of the 7th International Conference on Principles of Security and Trust (POST 2018).
 \end{center}

\begin{abstract}
Blockchain-based distributed computing platforms enable the trusted execution of computation---defined in the form of \emph{smart contracts}---without trusted agents.
Smart contracts are envisioned to have a variety of applications, ranging from financial to IoT asset tracking.
Unfortunately, the development of smart contracts has proven to be extremely error prone. In practice, contracts are riddled with security vulnerabilities comprising a critical issue since bugs are by design non-fixable and contracts may handle financial assets of significant value.
To facilitate the development of secure smart contracts, we have created the \emph{FSolidM} framework, which allows developers to define contracts as finite state machines (FSMs) with rigorous and clear semantics.
FSolidM provides an easy-to-use graphical editor for specifying FSMs, a code generator for creating Ethereum smart contracts, and a set of plugins that developers may add to their FSMs to enhance security and functionality.
\end{abstract}


\input{intro}

The rest of this paper is organized as follows.
In Section~\ref{sec:FSM}, we present blind auction as a motivating example, which we implement as an FSM-based smart contract. 
In Section~\ref{sec:tool}, we describe our FSolidM tool and its built-in plugins. 
Finally, in Section~\ref{sec:concl}, we offer concluding remarks and outline future work.






\section{Defining Smart Contracts as FSMs}
\label{sec:FSM}

Consider as an example a blind auction (similar to the one presented in \cite{solidityExample}), in which a bidder does not send her actual bid but only a hash of it (i.e., a blinded bid). A bidder is required to make a deposit---which does not need to be equal to her actual bid---to prevent her from not paying after she has won the auction. A deposit is considered valid if its value is higher than or equal to the actual bid. A blind auction has four main \textit{states}: 
\begin{compactenum}
\item \texttt{AcceptingBlindedBids}: blinded bids and deposits may be  submitted; 
\item \texttt{RevealingBids}:  bidders may reveal their bids (i.e., they can send their actual bids and the contract checks if the hash value is the same as the one submitted in the previous state and if they made sufficient deposit);
\item \texttt{Finished}: the highest bid wins the auction; bidders can withdraw their deposits except for the winner, who can withdraw only the difference between her deposit and bid;
\item \texttt{Canceled}: bidders can retract bids and withdraw their deposits.
\end{compactenum}

Since smart contracts have \textit{states} (e.g., \texttt{AcceptingBlindedBids}) and 
 provide functions that allow other entities (e.g., contracts or users) to invoke \textit{actions} that change the current state of a contract, they 
can be naturally represented as FSMs~\cite{solidityPatterns}. An FSM has a finite set of states and a finite set of transitions between these states. A transition forces a contract to take a set of actions if the associated conditions, i.e., the \textit{guards} of the transition, are satisfied. Since such states and transitions have intuitive meaning for developers, representing contracts as FSMs provides an adequate level of abstraction for behavior reasoning.

\begin{figure} [t]
\centering
\includegraphics[scale=0.8]{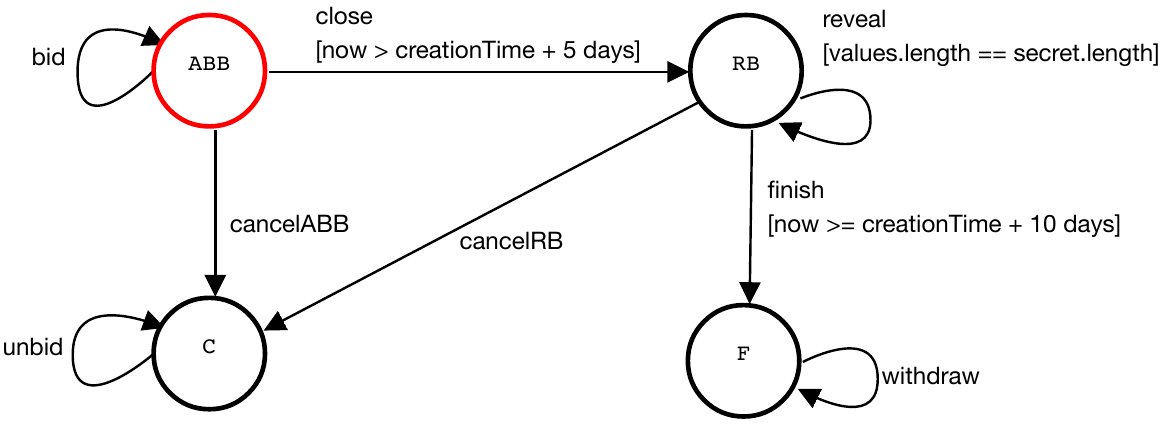}
\caption{Example FSM for blinded auctions.}
\label{fig:blauction}
\end{figure}

Figure \ref{fig:blauction} presents the blind auction example in the form of an FSM. For simplicity, we have abbreviated \texttt{AcceptingBlindedBids}, \texttt{RevealingBids}, \texttt{Finished}, and \texttt{Canceled} to \texttt{ABB}, \texttt{RB}, \texttt{F}, and \texttt{C}, respectively. \texttt{ABB} is the initial state of the FSM. Each transition (e.g., \texttt{bid}, \texttt{reveal}, \texttt{cancel}) is associated to a set of actions that a user can perform during the blind auction. For instance, a bidder can execute the \texttt{bid} transition at the \texttt{ABB} state to send a blind bid and a deposit value. Similarly, a user can execute the \texttt{close} transition, which signals the end of the bidding period, if the associated guard \texttt{now >= creationTime + 5 days} evaluates to true. To differentiate transition names from guards, we use square brackets for the latter. A bidder can reveal her bids by executing the \texttt{reveal} transition. The \texttt{finish} transition signals the completion of the auction, while the \texttt{cancelABB} and \texttt{cancelRB} transitions signal the cancellation of the auction. Finally, the \texttt{unbid} and \texttt{withdraw} transitions can be executed by the bidders to withdraw their deposits. 
For ease of presentation, we omit from Figure \ref{fig:blauction} the actions that correspond to each transition. For instance, during the execution of the \texttt{withdraw} transition, the following action is performed \texttt{amount = pendingReturns[msg.sender]}.  

Guards and actions are based on a set of variables (e.g., \texttt{creationTime}, \texttt{amount}). These variables represent data, which can be of type:
1)~\textit{contract data}, which is stored in the contract; 2)~\textit{input data}, which is received as transition input; 3)~\textit{output data}, which is returned as transition output.

\section{The FSolidM Tool}
\label{sec:tool}

FSolidM is an open-source\footnote{\url{https://github.com/anmavrid/smart-contracts}}, web-based tool that is built on top of WebGME~\cite{maroti2014next}. FSolidM enables collaboration between multiple users during the development of smart contracts. Changes in FSolidM are committed and versioned, which enables branching, merging, and viewing the history of a contract (Figure~\ref{fig:versioning}). 
We present the FSolidM tool in more detail in Appendix~\ref{sec:appendix}.

\begin{figure} [t]
\centering
\includegraphics[width=\textwidth]{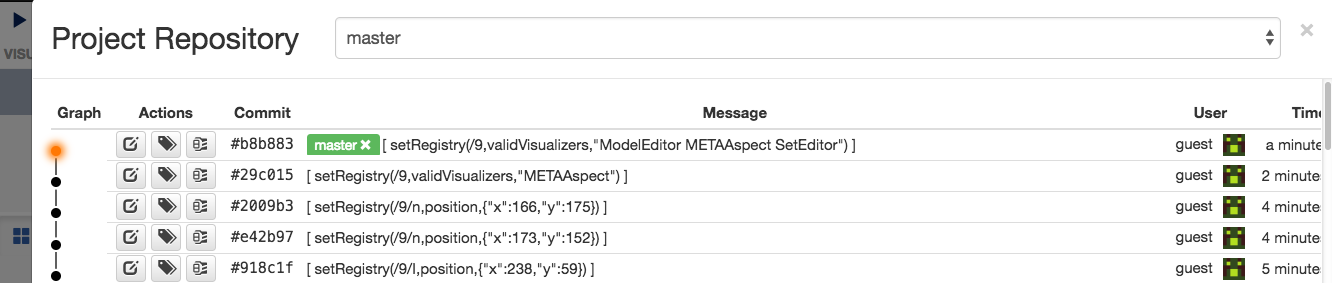}
\caption{Versioning in FSolidM.}
\label{fig:versioning}
\end{figure}

\begin{figure} [t]
\centering
\includegraphics[width=\textwidth]{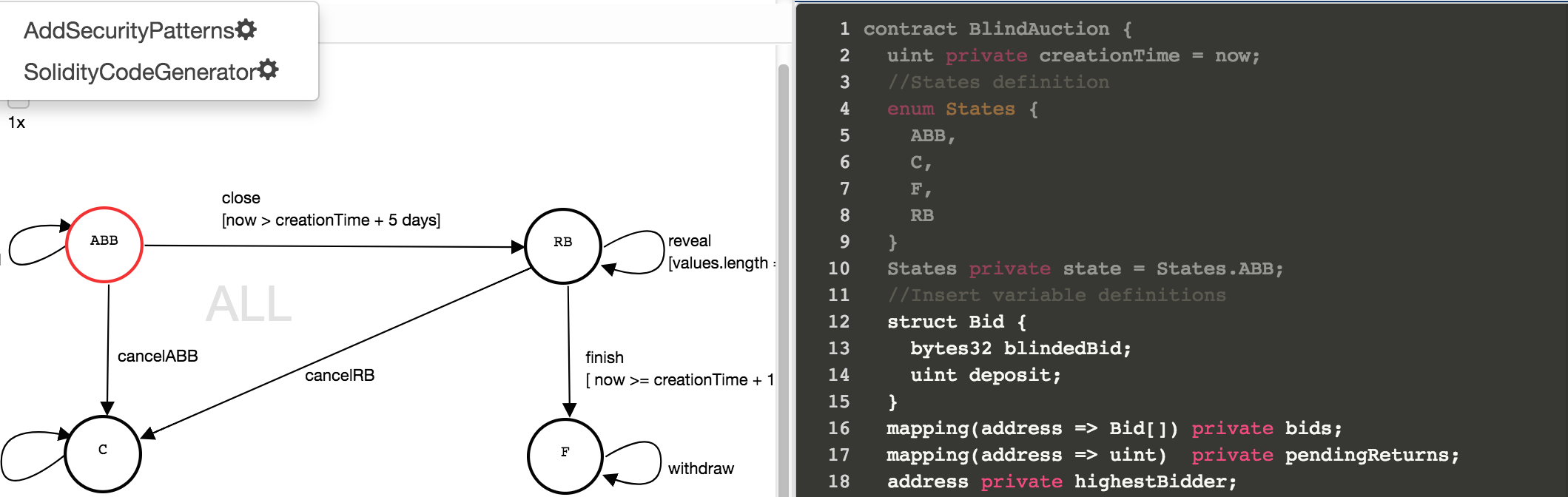}
\caption{The FSolidM model and code editors.}
\label{fig:code2}
\end{figure}

To generate the Solidity code of a smart contract using FSolidM, a user must follow three steps: 1)~specify the smart contract in the form of the FSM by using the dedicated graphical editor of FSolidM; 2)~specify attributes of the smart contract such as variable definition, statements, etc. in the \texttt{Property Editor} or in the dedicated Solidity code editor of FSolidM; 3) optionally apply security patterns and functionality extensions, and finally, generate the Solidity code. Figure \ref{fig:code2} shows the graphical and code editors of the tool (for steps 1 and 2) and the list of services (i.e., \texttt{AddSecurityPatterns} and \texttt{SolidityCodeGenerator} for step 3) that are provided by FSolidM. We have integrated a Solidity parser\footnote{\url{https://github.com/ConsenSys/solidity-parser}} to check the syntax of the Solidity code that is given as input by the users. 

Notice that in Figure \ref{fig:code2}, parts of the code shown in the code editor are darker (lines 1-10) than other parts (lines 12-15). The darker lines of code include code that was generated from the FSM model defined in the graphical editor and are locked---cannot be altered in the code editor. The non-dark parts indicate code that was directly specified in the code editor. 

FSolidM provides mechanisms for checking if the FSM is correctly specified (e.g., whether an initial state exists or not). FSolidM notifies developers of errors and provides links to the erroneous nodes of the model (e.g., a transition or a guard). Through the \texttt{SolidityCodeEditor} service, FSolidM provides an FSM-to-Solidity code generator. Additionally, through the \texttt{AddSecurityPatterns} service, 
%
%
%
%
%
\input{plugins.tex}

\input{concl}

\bibliographystyle{splncs}


\allowdisplaybreaks

\appendix










\section{Demonstration of FSolidM}
\label{sec:appendix}
We present in detail the FSolidM tool. In particular, we present how to design, apply security patterns, and generate the code of the \texttt{Blind Auction} contract. 

\begin{figure} [h]
\centering
\includegraphics[width=\textwidth]{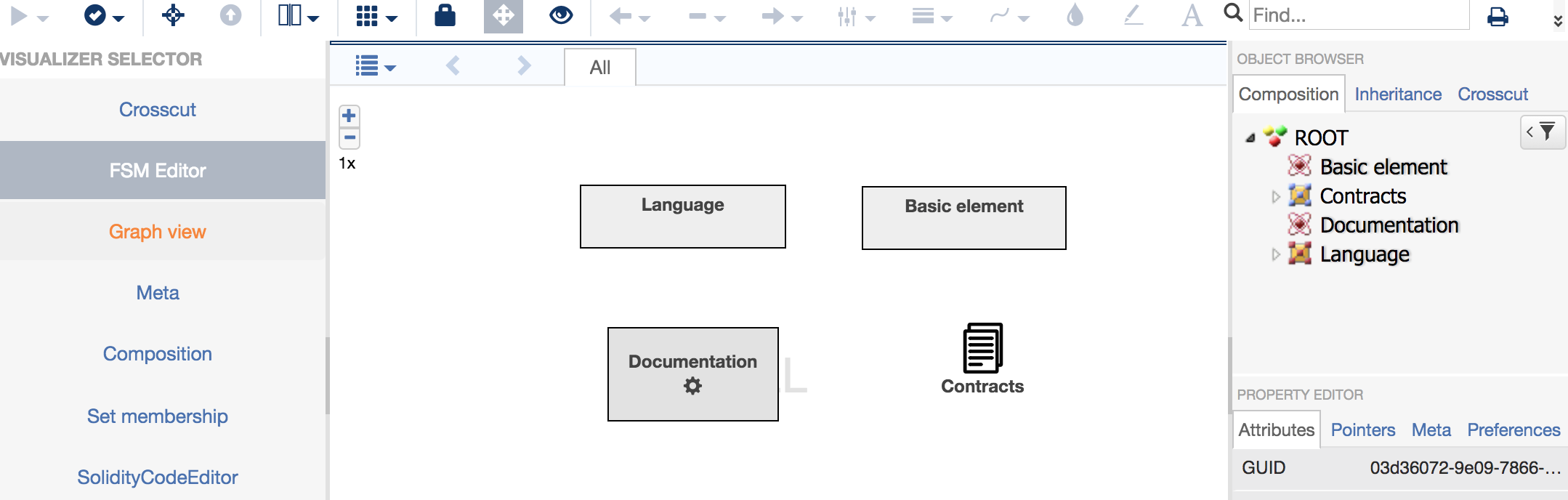}
\caption{FSolidM initial page.}
\label{fig:First}
\end{figure}

\subsection{The FSolidM Environment}
\label{sec:fsolidm_environment}

FSolidM is a web-based, open-source\footnote{\url{https://github.com/anmavrid/smart-contracts}} tool. The initial page of FSolidM is shown in Figure \ref{fig:First}. The user can navigate to the Contracts page by double-clicking the \texttt{Contracts} icon. 

\begin{figure} 
\centering
\includegraphics[width=\textwidth]{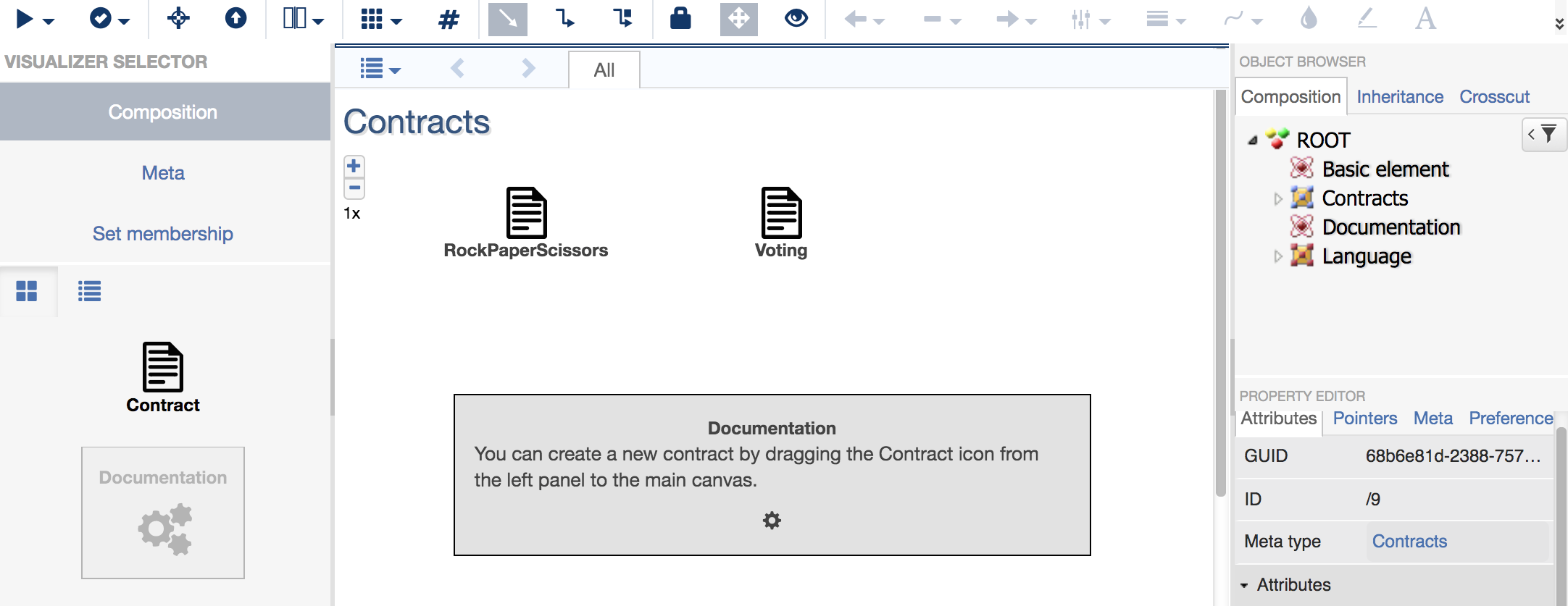}
\caption{Contracts page.}
\label{fig:Second}
\end{figure}

The \texttt{Contracts} page is shown in Figure \ref{fig:Second}. 
The center panel is the Canvas. 
The left side of the screen shows the \texttt{Visualizer Selector} options and below them the \texttt{Part Browser}, which displays the concepts that can be instantiated inside the Canvas. For instance, dragging and dropping a \texttt{Contract} from the \texttt{Part Browser} creates a new contract on the Canvas. 
The top right corner of the user interface is the \texttt{Object Browser}, which shows the composition hierarchy of FSolidM starting at Root. Embeddable documentation can be added at every level of this hierarchy. For instance, in Figure \ref{fig:Second}, we have added documentation explaining how to create a New contract. Below the \texttt{Object Browser} is the \texttt{Property Editor}, where attributes, preferences, and other properties of the currently selected contract can be edited.

\begin{figure} [t]
\centering
\includegraphics[width=\textwidth]{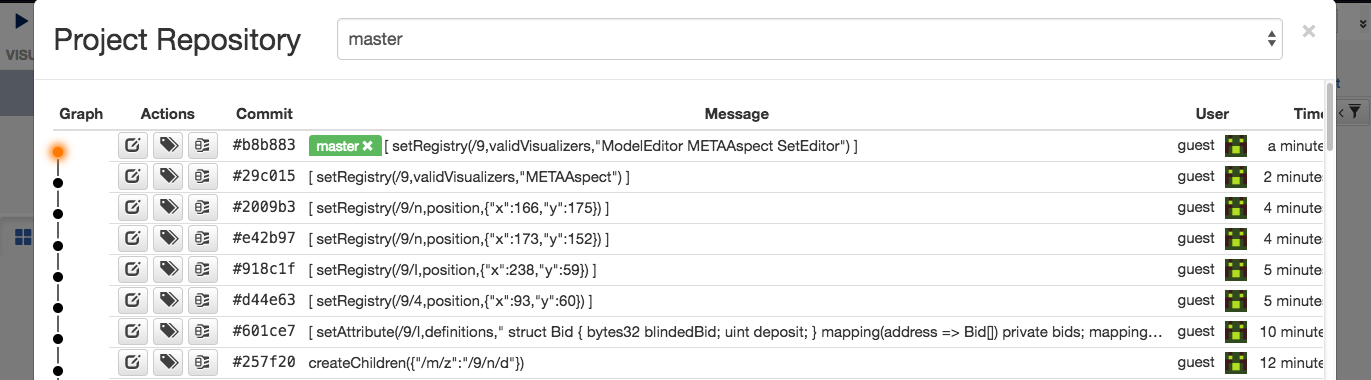}
\caption{Versioning in FSolidM.}
\label{fig:version}
\end{figure}

\begin{figure}
\centering
\includegraphics[width=\textwidth]{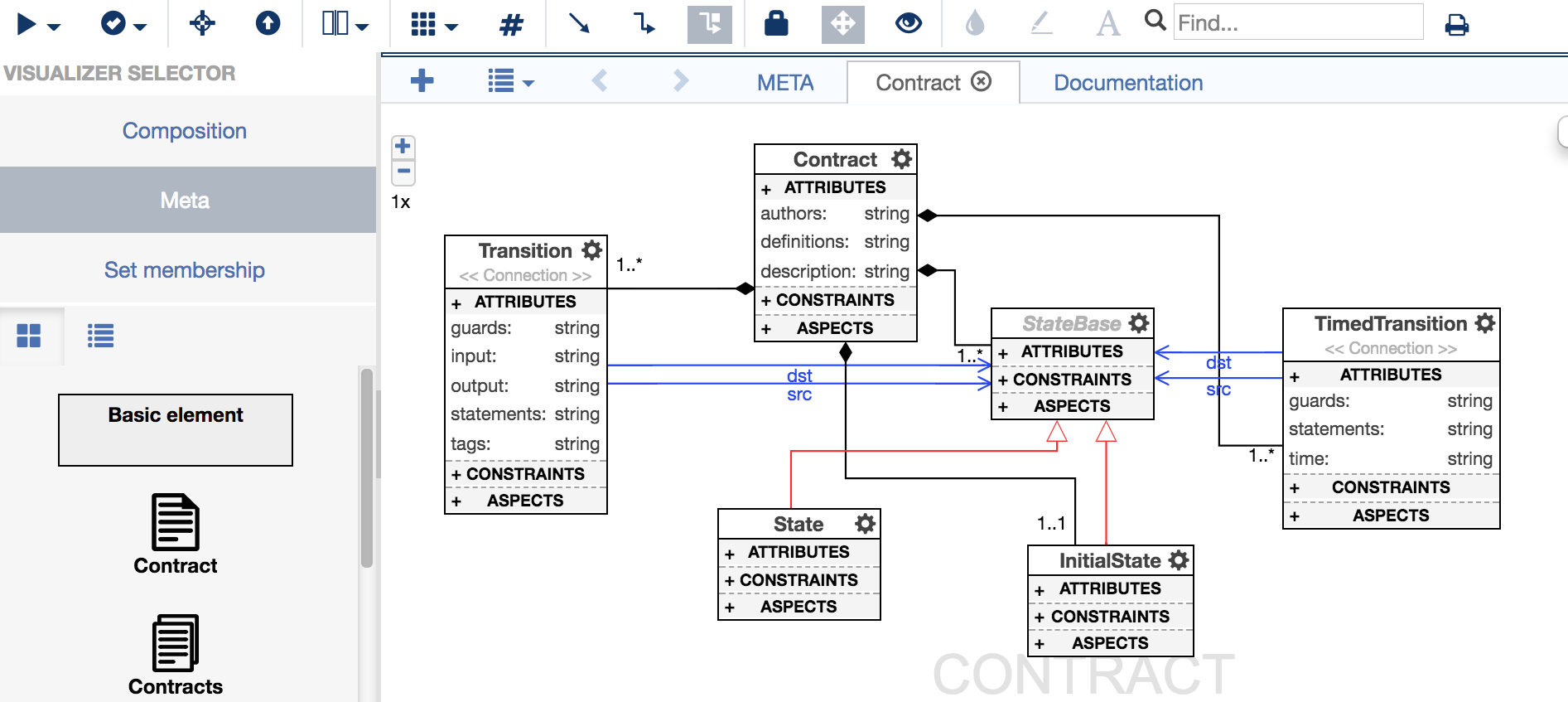}
\caption{The FSolidM language.}
\label{fig:meta}
\end{figure}

FSolidM provides a collaborative and versioned environment. Multiple users may work on the same smart contract simultaneously. Changes are immediately broadcasted to all users and everyone sees the same state. This is similar to how Google Docs works. 
Changes in FSolidM are committed and versioned, which enables branching, merging, and viewing the history of a contract. Figure \ref{fig:version} shows the history of the \texttt{master} branch. 


\subsection{The FSolidM Language}


Next, we present the main elements of our FSM-based smart-contract language, which is defined in FSolidM as a UML class diagram (Figure \ref{fig:meta}). By double-clicking \texttt{Meta} in the \texttt{Visualizer Selector} shown in Figure \ref{fig:Second}, we navigate to the language-specification of FSolidM. The behavior of a \texttt{Contract} is described as an FSM. The \texttt{State\_Base} element is abstract and it can be instantiated by either an \texttt{InitialState} or a \texttt{State}. Each contract must have exactly one \texttt{InitialState}, which is enforced by the cardinality of the containment relation. The \texttt{Transition} element is characterized by six attributes: 1)~\texttt{name}, which is inherited from \texttt{Basic Element} of our language, 2)~the associated \texttt{guards}, the 3)~\texttt{input} and 4)~\texttt{output} data of the transition, the 5)~\texttt{statements}, and finally, the 6)~\texttt{tags}. Similarly, the \texttt{Timed Transition} element is characterized by three attributes: 1)~\texttt{guards}, 2)~\texttt{statements}, and 3)~\texttt{time}.


\subsection{Designing a Blind Auction Smart Contract}

\begin{wrapfigure}{r}{0.65\textwidth}
\centering
\includegraphics[width=0.65\textwidth]{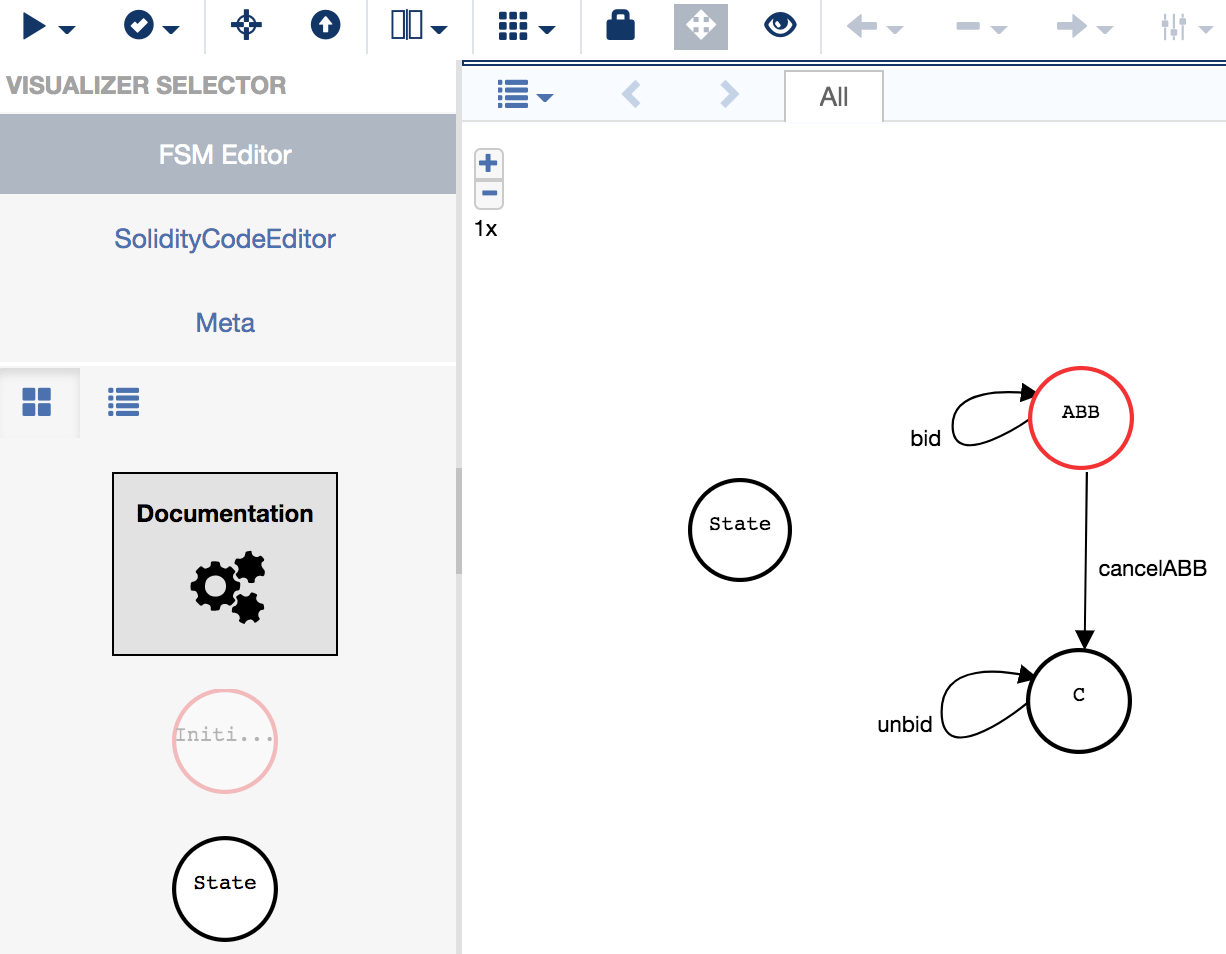}

\caption{Designing a blind auction smart contract.}
\label{fig:contract}
\end{wrapfigure}

Next, we present how to design a blind auction contract as an FSM using FSolidM. After creating a new \texttt{Contract} on the canvas shown in Figure \ref{fig:Second} (as described in Section~\ref{sec:fsolidm_environment}), the user must  specify a unique name for this contract, e.g., \texttt{BlindAuction}, in the \texttt{name} attribute of the Contract listed in the \texttt{Property Editor}. Then, the user creates the states of the FSM (Figure \ref{fig:contract}) by drag-and-dropping states from the \texttt{Part Browser}. By clicking on a state, the user may add a transition to another state or the same state. For each transition, the user may specify its corresponding attributes, e.g., the guards, inputs, and statements at the \texttt{Property Editor} or at the dedicated Solidity code editor that we have developed. 
The user may navigate to the code editor by double-clicking \texttt{SolidityCodeEditor} in the \texttt{Visualizer Selector}. 

\begin{figure}
\centering
\includegraphics[width=\textwidth]{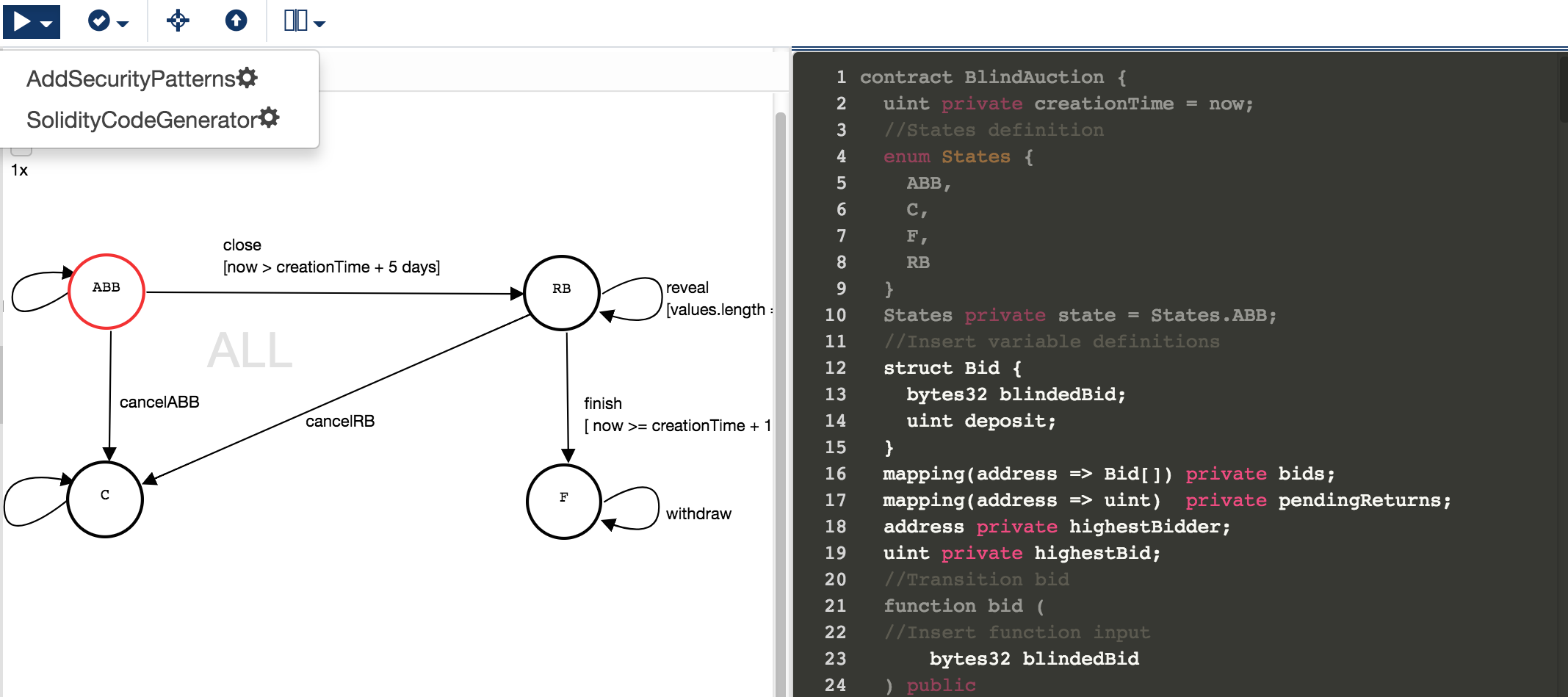}
\caption{The FSolidM model and code editors.}
\label{fig:code}
\end{figure}

Figure~\ref{fig:code} shows the model and code editors and the equivalent representation of the blind auction contract in the two editors. From the FSM shown in the model editor, we generate equivalent Solidity code, which consists in the darker parts of the code shown in the code editor. This part of the code is generated automatically from the FSM and cannot be altered by the user in the code editor. The user can only change the FSM in the model editor. The lighter parts of the code in the code editor represent parts of the code that the used has directly defined in the code editor. For instance, the user may specify variables, transition statements and guards, directly in the code editor. The code and model editors are tightly integrated. Once a user makes a change in the model editor, the code editor is updated and vice versa. 

\subsection{Applying Security Patterns}

\begin{figure}
\centering
\includegraphics[width=\textwidth]{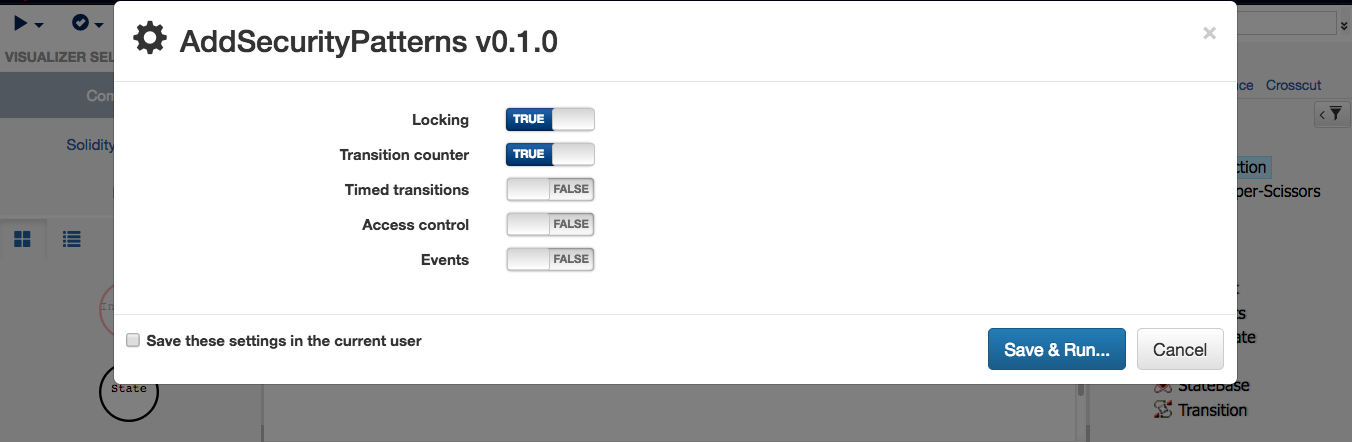}
\caption{Running the \texttt{AddSecurityPatterns}.}
\label{fig:patterns}
\end{figure}

To apply security patterns and functionality extensions, a user must double-click the \texttt{AddSecurityPatterns} service offered by the dropped down menu in the upper left corner of the tool (Figure~\ref{fig:code}). Then, the widget shown in Figure~\ref{fig:patterns} pops up, and the developer may pick the patterns to be applied.

FSolidM plugins not only enhance security but also increase the computational cost of transitions. Since users must pay a relatively high price for computation performed on the public Ethereum platform, the computational cost of plugins is a critical question. We measured and compared the computational cost of transitions in our blind auction contract without and with the locking and transition counter plugins by using
Solidity compiler version 0.4.17 with optimizations enabled.

\begin{figure}
\centering
\begin{tikzpicture}
\begin{axis}[
    width=0.9\textwidth,
    height=0.5\textwidth,
    bar width=0.018\textwidth,
    ybar,
    ymin=0,
    ylabel={Transaction cost [gas]},
    legend pos=north east,
    symbolic x coords={{bid}, {cancelABB}, {unbid}, {close}, {reveal}, {finish}, {withdraw}},
    xtick=data,
    xticklabel style={font=\small\tt},
    ]
    \addplot coordinates { 
        (bid, 58249)        
        (cancelABB, 42059)  
        (unbid, 19735)
        (close, 42162)
        (reveal, 65729)
        (finish, 27239)
        (withdraw, 20290)
    };
    \addlegendentry{without plugins};
    \addplot coordinates { 
        (bid, 68917)        
        (cancelABB, 52727)
        (unbid, 30406)
        (close, 52830)
        (reveal, 76415)
        (finish, 37913)
        (withdraw, 30961)
    };
    \addlegendentry{with locking};
    \addplot coordinates { 
        (bid, 63924)        
        (cancelABB, 47661)
        (unbid, 25406)
        (close, 47764)
        (reveal, 71390)
        (finish, 32891)
        (withdraw, 25961)
    };
    \addlegendentry{with counter};
    \addplot coordinates { 
        (bid, 74607)        
        (cancelABB, 58329)
        (unbid, 36074)
        (close, 58432)
        (reveal, 82067)
        (finish, 43559)
        (withdraw, 36629)
    };
    \addlegendentry{with both};
\end{axis}
\end{tikzpicture}
\caption{Transaction costs in gas without plugins (\textcolor{blue}{\bf blue}), with locking plugin (\textcolor{red}{\bf red}), with transition counter plugin (\textcolor{brown}{\bf brown}), and with both plugins (\textcolor{darkgray}{\bf dark gray}).}
\label{fig:transCosts}
\end{figure}
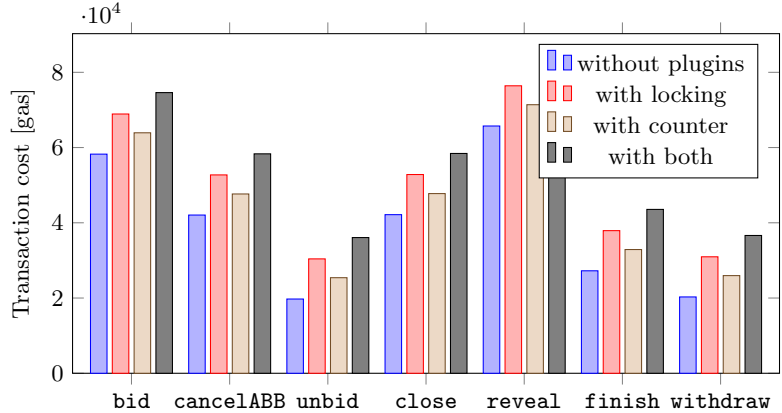

We quantified computational cost of a transition as the gas cost of an Ethereum transaction that invokes the function implementing the transition. Gas measures the cost of executing computation on the Ethereum platform.
The cost of deploying our smart contract was 504,672 gas without any plugins, 577,514 gas with locking, 562,800 gas with transition counter, and 637,518 gas with both.
Figure~\ref{fig:transCosts} shows the gas cost of each transition for all four combinations of the two plugins. 
We make two key observations.
First, \emph{computational overhead is almost constant} for both plugins and also for their combination.
Second, the \emph{computational overhead of the two plugins is additive}.

\subsection{Generating the Solidity Code}

\begin{figure}
\centering
\includegraphics[width=\textwidth]{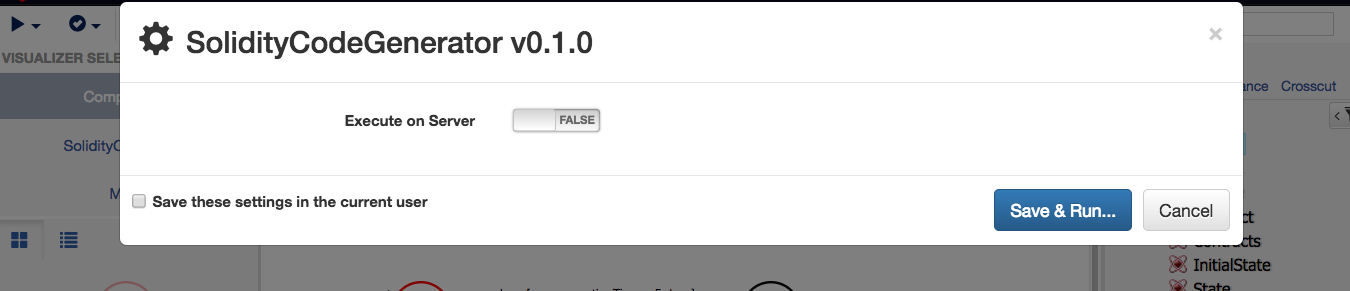}
\caption{Running the \texttt{SolidityCodeGenerator}.}
\label{fig:generator}
\end{figure}

To generate Solidity code a user must on the \texttt{SolidityCodeGenerator} service offered by the dropped down menu in the upper left corner of the tool (Figure \ref{fig:code}). Then, the widget shown in Figure \ref{fig:generator} pops up, and the developer can click on the \texttt{Save \& Run} button to continue with the code generation. 

\begin{figure}
\centering
\includegraphics[width=\textwidth]{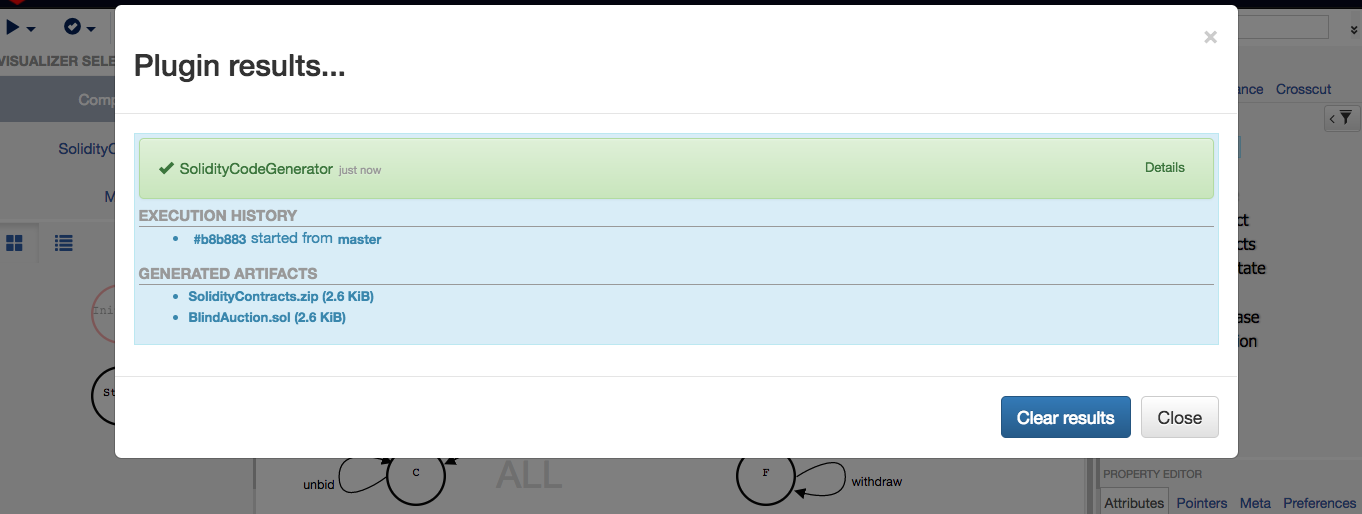}
\caption{Successful Solidity code generation.}
\label{fig:correct}
\end{figure}

\begin{figure}
\centering
\includegraphics[width=\textwidth]{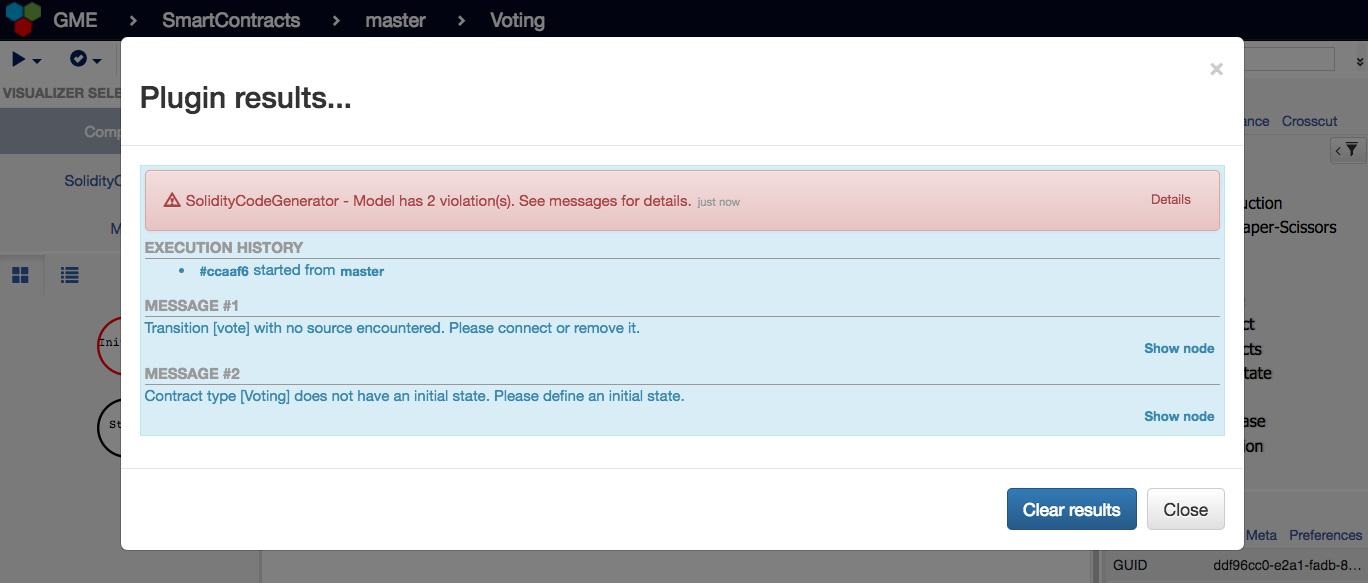}
\caption{Unsuccessful Solidity code generation due to incorrect input.}
\label{fig:wrong}
\end{figure}

If the code generation is successful, i.e., there are no specification errors in the given input, then the widget shown in Figure \ref{fig:correct} pops-up. The developer can then click on the generated artifacts to download the generated Solidity contract. If, on the other hand, the code generation is not successful due to incorrect input, then, the widget shown in Figure \ref{fig:wrong} pops-up. As shown in Figure \ref{fig:wrong}, FSolidM lists the errors found in the specification of the contract with detailed explanatory messages. Additionally, FSolidM provides links (through \texttt{Show node}) that redirect the developer to the erroneous nodes of the contract.


\end{document}

%% file: intro.tex
\section{Introduction}
\label{sec:intro}

In recent years, blockchains have seen wide adoption.
For instance, the market capitalization of Bitcoin, the leading blockchain-based cryptocurrency, has grown from \$15 billion to more than \$100 billion in 2017.
The goal of the first generation of blockchains was only to provide cryptocurrencies and payment systems.
In contrast, more recent blockchains, such as Ethereum, strive to provide distributed computing platforms~\cite{underwood2016blockchain,wood2014ethereum}.
Blockchain-based distributed computing platforms enable the trusted execution of general purpose computation, implemented in the form of \emph{smart contracts}, without any trusted parties.
Blockchains and smart contracts are envisioned to have a variety of applications, ranging from finance to IoT asset tracking~\cite{christidis2016blockchains}.
As a result, they 
are embraced by an increasing number of organizations and companies, including
major IT and financial firms, such as Cisco, IBM, Wells Fargo, and J.P. Morgan~\cite{vukolic2017rethinking}.

However, the development of smart contracts has proven to be extremely error prone in practice. 
Recently, an automated analysis of a large sample of smart contracts from the Ethereum blockchain found that more than 43\% of contracts have security issues~\cite{luu2016making}.
These issues often result in security vulnerabilities, which may be exploited by cyber-criminals to steal cryptocurrencies and other digital assets.
For instance, in 2016, \$50 million worth of cryptocurrencies were stolen in the infamous ``The DAO'' attack, which exploited a combination of smart-contract vulnerabilities~\cite{finley2016million}.
In addition to theft, malicious attackers may also be able to cause damage by leading a smart contract into a deadlock, which prevents account holders from spending or withdrawing their own assets.

The prevalence of smart-contract vulnerabilities poses a severe problem in practice due to multiple reasons. 
First, smart contracts handle assets of significant financial value: at the time of writing, contracts deployed on the Ethereum blockchain together hold more than \$6 billion worth of cryptocurrency. 
Second, it is \emph{by design} impossible to fix bugs in a contract (or change its functionality in any way) once the contract has been deployed. 
Third, due to the ``code is law'' principle~\cite{bhargavan2016short}, it is also \emph{by design} impossible to remove a faulty or malicious transaction from the blockchain, which means that it is often impossible to recover from a security incident.\footnote{It is possible to remove a transaction or hard fork the blockchain if the stakeholders reach a consensus; however, this undermines the trustworthiness of the platform~\cite{leising2017ether}.} 

Previous work focused on alleviating security issues in \emph{existing} smart contracts by providing tools for verifying correctness~\cite{bhargavan2016short} and for identifying common vulnerabilities~\cite{luu2016making}.
In contrast, we take a different approach by developing a framework, called \emph{FSolidM}~\cite{fsolidm}, which helps developers to create smart contracts that are secure by design.
The main features of our framework are as follows.

\noindent \textbf{Formal Model}: 
One of the key factors contributing to the prevalence of security issues is the semantic gap between the developers' assumptions about the underlying execution semantics and the actual semantics of smart contracts~\cite{luu2016making}.
To close this semantic gap, FSolidM is based on a simple, formal, finite-state machine (FSM) based model for smart contracts, which we introduced in~\cite{fsolidm}.
The model was designed to support Ethereum smart contracts, but it could easily be extended to other platforms.

\noindent \textbf{Graphical Editor}:
To further decrease the semantic gap and facilitate development, FSolidM provides an easy-to-use graphical editor that enables developers to design smart contracts as FSMs.

\noindent \textbf{Code Generator}:
FSolidM provides a tool for translating FSMs into Solidity, the most widely used high-level language for developing Ethereum contracts. Solidity code can be translated into Ethereum Virtual Machine bytecode, which can be deployed and executed on the platform.

\noindent \textbf{Plugins}:
FSolidM enables extending the functionality of FSM based smart contract using plugins.
As part of our framework, we provide a set of plugins that address common security issues and implement common design patterns, which were identified by prior work~\cite{luu2016making,atzei2017survey,bartoletti2017empirical}.
In Table~\ref{tab:common}, we list these vulnerabilities and patterns with the corresponding plugins. 

\begin{table}[t]
\setlength{\tabcolsep}{0.75em}
\renewcommand{\arraystretch}{1.0}
\caption{Common Smart-Contract Vulnerabilities and Design Patterns}
\label{tab:common}
\centering
\begin{tabular}{|l|l|l|}
\hline
Type & Common Name & FSolidM Plugin \\
\hline\hline
\multirow{2}{*}{Vulnerabilities} & reentrancy~\cite{luu2016making,atzei2017survey} & locking \\ 
\cline{2-3}
& transaction ordering~\cite{luu2016making,atzei2017survey} & transition counter \\ 
\hline
\multirow{2}{*}{Patterns} & time constraint~\cite{bartoletti2017empirical} & timed transitions \\ 
\cline{2-3}
& authorization~\cite{bartoletti2017empirical} & access control \\ 
\hline
\end{tabular}
\end{table}

\noindent \textbf{Open Source}:
FSolidM is open-source and available online (see Section~\ref{sec:tool}).

The advantages of our framework, which helps developers to create secure contracts instead of trying to fix existing ones, are threefold.
First, we decrease the semantic gap and eliminate the issues arising from it by providing a formal model and an easy-to-use graphical editor. 
Second, since the process is rooted in rigorous semantics, our framework may be connected to formal analysis tools~\cite{bensalem2009d,cavada2014nuxmv}.
Third, the code generator and plugins enable developers to implement smart contracts with minimal amount of error-prone manual coding.

%% file: plugins.tex
\begin{figure}
\centering
\includegraphics[width=\textwidth]{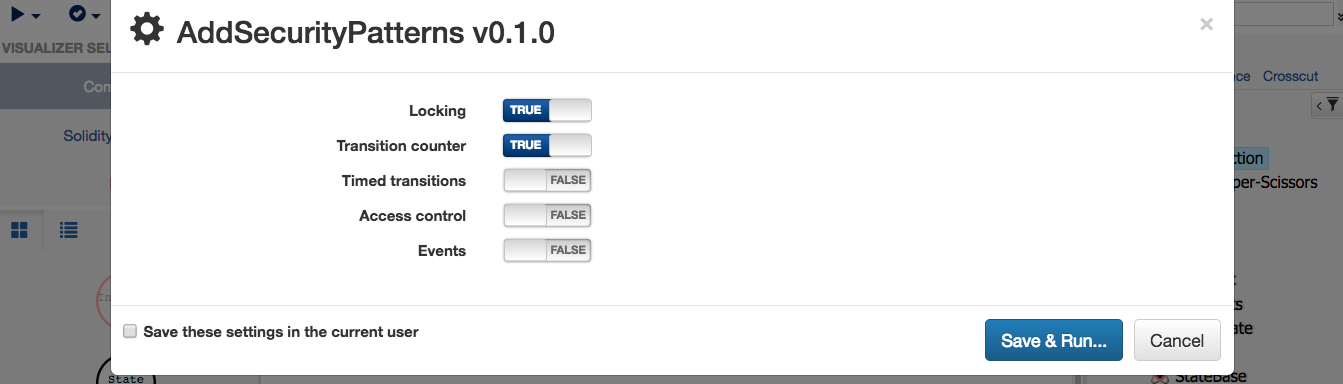}
\caption{Running the \texttt{AddSecurityPatterns}.}
\label{fig:plugins}
\end{figure}

%
FSolidM enables developers to enhance the functionality and security of contracts conveniently by adding plugins to them. Our framework provides four built-in plugins: locking, transition counter, timed transitions, and access control. Plugins can be simply added with a ``click,'' as shown in Figure~\ref{fig:plugins}.

\subsubsection{Locking}
\label{sec:locking}
When an Ethereum contract calls a function of another contract, the caller has to wait for the call to finish. 
This allows the callee---who may be malicious---to exploit the intermediate state of the caller, e.g., by invoking a function of the caller.
This re-entrancy issue is one of the most well-known vulnerabilities, which was also exploited in the infamous ``The DAO'' attack.

To prevent re-entrancy, we provide a security plugin for locking the smart contract.
Locking eliminates re-entrancy vulnerabilities in a ``foolproof'' manner: 
functions within the contract cannot be nested within each other in any way.

\subsubsection{Transition Counter}
\label{sec:transCounter}

If multiple functions calls are invoked around the same time, then the order in which these calls are executed on the Ethereum blockchain may be unpredictable.
Hence, when a user invokes a function, she may be unable to predict what the state and the values stored within a contract will be when the function is actually executed. 
This issue has been referred to as ``transaction-ordering dependence''~\cite{luu2016making} and ``unpredictable state''~\cite{atzei2017survey}, and it can lead to various security vulnerabilities.

We provide a plugin that prevents unpredictable-state vulnerabilities by enforcing a strict ordering on function call executions.
The plugin expects a transition number in every function as a parameter and ensures that the number is incremented by one for each function execution.
As a result, when a user invokes a function with the next transition number in sequence, she can be sure that the function is executed before any other state changes can take place.

\subsubsection{Automatic Timed Transitions}
\label{sec:timedTrans}

We provide a plugin for implementing time-constraint patterns. We extend our language with timed transitions,
which are similar to non-timed transitions, but 1)~their guards and assignments do not use input or output data and 2)~they include a number specifying transition time.

We implement timed transitions as a modifier that is applied to every function, and which ensures that timed transitions are executed automatically if their time and data guards are satisfied. 
%
Writing such modifiers manually could lead to vulnerabilities.
For example, a developer might forget to add a modifier to a function, which enables malicious users to invoke functions without the contract progressing to the correct state (e.g., place bids in an auction even though the auction should have already been closed due to a time limit).

\subsubsection{Access Control}
\label{sec:accessContr}

In many contracts, access to certain transitions (i.e., functions) needs to be controlled and restricted.  
For example, any user can participate in a typical blind auction by submitting a bid, but only the creator should be able to cancel the auction. 
To facilitate the enforcement of such constraints, we provide a plugin that 1) manages a list of administrators at runtime (identified by their addresses) and 2) enables developers to forbid non-administrators from accessing certain functions.

%% file: concl.tex
\section{Conclusion and Future Work}
\label{sec:concl}

Blockchain-based decentralized computing platforms with smart-contract functionality are envisioned to have a significant technological and economic impact in the future.
However, if we are to avoid an equally significant risk of security incidents,
we must ensure that smart contracts are secure.
%
To facilitate the development of smart contracts that are secure by design, we created the FSolidM framework, which  enables designing contracts as FSMs.
Our framework is rooted in rigorous yet clear semantics, and it provides an easy-to-use graphical editor and code generator.
We also implemented a set of plugins that developers can use to enhance the security or functionality of their contracts. 
%
In the future, we plan
%
to integrate model checkers and compositional verification tools into our framework~\cite{bensalem2009d,cavada2014nuxmv} to enable the verification of  security and safety properties.
%